\documentclass[pra,twocolumn,tightenlines,nofootinbib]{revtex4}
\usepackage{graphicx}
\usepackage{multirow}
\usepackage{amsmath,amsfonts,amssymb}
\usepackage{bm}
\usepackage{dcolumn}
\usepackage{graphicx}
\usepackage{color, soul}

\newcommand{\eref}[1]{Eq.~(\ref{#1})}
\newcommand{\tref}[1]{Table~\ref{#1}}

\begin{document}

%\title{An accurate calculation of $M1$ and $E2$ polarizabilities and hyperpolarizabilities
%of the clock states in divalent atoms}
\title{Contribution of negative-energy states to multipolar polarizabilities of the Sr optical lattice clock}

\author{S.~G.~Porsev$^{1}$}
\author{M.~G.~Kozlov$^{2,3}$}
\author{M.~S.~Safronova$^{1,4}$}

\affiliation{
$^1$Department of Physics and Astronomy, University of Delaware, Newark, Delaware 19716, USA\\
$^2$Petersburg Nuclear Physics Institute of NRC ``Kurchatov Institute'', Gatchina, Leningrad District 188300, Russia\\
$^3$St.~Petersburg Electrotechnical University ``LETI'', Prof. Popov Str. 5, St.~Petersburg 197376,  Russia \\
$^4$Joint Quantum Institute, National Institute of Standards and Technology and the University of Maryland, Gaithersburg, Maryland 20742, USA}
\date{\today}

\begin{abstract}
We address the problem of lattice light shifts in the Sr clock caused by multipolar $M1$ and $E2$ atom-field interactions. We presented a simple but accurate
formula for the magnetic-dipole polarizability that takes into account both the positive and negative energy states contributions. We calculated the contribution of negative energy states to the $M1$ polarizabilities of the clock
$^1\!S_0$ and $^3\!P_0^o$ states at the magic frequency. Taking these contributions into account, we obtained good agreement with the experimental results, resolving the major discrepancy between the theory and the experiment.
\end{abstract}
\pacs{ }

\maketitle
%=======================
\paragraph{Introduction}
%=======================
The past decade brought forth extraordinary improvements in the accuracy and stability of atomic Sr optical clocks based on the $^1\!S_0 -\, ^3\!P_0^o$ transition. In 2015, the systematic uncertainty of the optical lattice clock of Sr was reported to be $2.1\times 10^{-18}$ in fractional frequency units ~\cite{NicCamHut15}.   In 2022, the resolution of the gravitational redshift across a millimeter-scale atomic sample was demonstrated \cite{2022JunSr}, as well as record stability, reaching $10^{-18}$ in just a few seconds.
%In recent years the measurement precision was improved by an order of magnitude and a fractional frequency uncertainty was reduced to a mid of $10^{-19}$~\cite{KimAepBot23}.
Improved clock precision is needed for many fundamental and practical applications, including relativistic geodesy \cite{TanKat21},  search for the variation of fundamental constants \cite{Lange_2021}, and dark matter \cite{Arv15,2020SrcavityDM,roberts_search_2020,filzinger2023improved}, tests of general relativity \cite{2020Skytree,FOCOS}, searches for violation of Lorentz invariance\cite{2019YbLLI}, redefinition of the second \cite{BreMilPiz17}, detection of gravitational waves \cite{2016clockGW,Fedderke_2022}, and others. Reaching $10^{-19}$ and better uncertainty with optical lattice clocks requires a further understanding of systematic light shifts caused by the trapping laser creating the optical lattice.

When an atom is placed in a laser field, atomic energy levels experience a shift
due to the interaction of the atom with the electromagnetic field of the laser wave.
The dominant part of this shift is proportional to the laser intensity and is determined by the difference of the electric dipole (E1) polarizabilities of two clock
states~\cite{MitSafCla10} at the wavelength of the trapping laser. %\MK{trapping laser}. 
To cancel out this shift, the laser wavelength is chosen so that E1 polarizabilities of the clock levels are the same, so the atom experiences the same Stark shift in both states. If the trapping laser of the optical lattice clock operates at such a ``magic'' wavelength~\cite{KatIdoKuw99,YeVerKim99}, the dominant light shift of the clock states cancels out in the clock transition.

This cancellation is not complete because there are other contributions to the light shift, caused by the magnetic dipole ($M1$) and electric-quadrupole ($E2$) interactions of the atom with the lattice field and determined by the $M1$ and $E2$ polarizabilities of the clock state, as well as hyperpolarizability \cite{PorSafSaf18}. When the systematic uncertainties of the clock reached $10^{-18}$, this effect became significant and required further study~\cite{OvsPalTai13,KatOvsMar15,BroPhiBel17,UshTakKat18,DorKloPal23,KimAepBot23}. %\textbf{[MK: Ref.8 is updated]}.

Calculation of the quantity $\Delta \alpha_{qm} \equiv \Delta \alpha_{M1} + \Delta\alpha_{E2}$ at the magic wavelength $\lambda^* = 813.4280(5)$ nm~\cite{YeKimKat08}, where
\begin{eqnarray}
\Delta \alpha_{M1} &\equiv& \alpha_{M1}(^3\!P^o_0) - \alpha_{M1}(^1\!S_0),  \notag \\
\Delta \alpha_{E2} &\equiv& \alpha_{E2}(^3\!P^o_0) - \alpha_{E2}(^1\!S_0),
\label{delEM}
\end{eqnarray}
%$\Delta \alpha_{M1}$ and $\Delta \alpha_{E2}$ are the differential polarizabilities of the $^3\!P_0^o$ and $^1\!S_0$ states,
was performed in Refs.~\cite{PorSafSaf18,FanTanYon19}. Although the theoretical results were in good agreement with each other, they differed even in sign from the experimental results \cite{UshTakKat18,DorKloPal23,KimAepBot23}.

An explanation of this discrepancy was suggested in the recent paper~\cite{WuShiTan23}, which included the contribution of negative energy intermediate states in calculating the polarizabilities $M1$ and $E2$ of
the clock states at the magic frequency, which was not considered in Refs.~\cite{PorSafSaf18,FanTanYon19}. However, the precision of the calculation was around 50\%, which was insufficient to differentiate between the experimental measurements. The paper also omitted a rather large contribution of the core electrons. The accuracy of the method that was used in \cite{WuShiTan23,WuShiNi23}, that is, the direct inclusion of negative energy states in all numerical parts of the calculation, is difficult to significantly improve. It is also difficult to directly include negative energies in the calculation of polarizabilities with more accurate approaches, such as the CI+all-order method that combined configuration interaction and coupled cluster approaches \cite{PorSafSaf18}. This is due to the complexity of modifying a very large suite of codes to include negative energies in every step of both the CI and the coupled-cluster computations.  Meanwhile,  reliable theoretical calculations of multipolar polarizability for Sr and other atoms used in optical lattice clocks are urgently needed, especially due to some disagreement between experimental results \cite{UshTakKat18,KimAepBot23}.

In this work, we derive an analytical formula for the contribution of negative-energy states to magnetic dipole polarizability that only needs a numerical computation of a single matrix element, thus resolving the major problem of accurate theoretical computation of multipolar polarizabilities.   We evaluate the accuracy of the new approach and use it to compute the multipolar polarities of the Sr clock. We also present an explanation of why negative energy contributions happen to be so important for the M1 polarizabilities while negligible for the E2 polarizabilities.
%============================
\paragraph{General formalism.}
%============================
We assume that an atom in a state $|0\rangle$ with $J=0$ is placed in a linearly polarized field of the lattice standing wave with the electric field vector along the $z$-axis, given by
\begin{equation}
{\mathcal E}_z = 2 {(\mathcal E_0)}_z \, {\rm cos}(kx)\, {\rm cos}(\omega t) ,
\end{equation}
where $k=\omega/c$, $\omega$ is the lattice laser wave frequency and $c$ is the speed of light.

The optical lattice potential for the atom at $|kx| \ll 1$, where $x$ determines the position of the atom starting from the standing wave antinode, can be approximated up to terms $\sim \mathcal{E}_0^2$ as~\cite{OvsPalTai13,KatOvsMar15}
\begin{eqnarray}
U(\omega) \approx &-& \alpha_{E1}(\omega)(1-k^2x^2)\,\mathcal{E}_0^2  \notag \\
                  &-& \{\alpha_{M1}(\omega) + \alpha_{E2}(\omega)\}k^2 x^2\,\mathcal{E}_0^2 .
\label{DeltaE}
\end{eqnarray}
%\textbf{MK: Naively one may think that E2 operator interacts with grad E. Then, for this operator, the nodes become antinodes and vice versa. I guess this is not true, but still, are we sure that this equation is correct?}
%\SP{1) This equation was derived in [10,11] 10 years ago (we just refer to it). I guess it was checked repeatedly by the different experimental groups. 2) From this equation we only need to know that it depends on $\alpha_{M1}$ and $\alpha_{E2}$, all the rest is not important for our work. Not sure that we need to spend time to rederive this equation}.

The ac $2^K$-pole polarizability of the $|0\rangle$ state can be expressed (if not stated otherwise, we use atomic units $\hbar=m=|e|=1,\,c \approx 137$) as~\cite{PorDerFor04}
\begin{eqnarray}
\alpha_{\lambda K}(\omega) &=& \frac{K+1}{K\,[(2K-1)!!]^2} \left(\frac{\omega}{c}\right)^{2K-2}  \notag \\
&\times& \sum_n \frac{(E_n-E_0) | \langle n|(T_{\lambda K})_0 |0 \rangle |^2}{(E_n-E_0)^2-\omega^2} .
\label{Qlk}
\end{eqnarray}
Here, $\lambda$ distinguishes between electric, $\lambda = E$, and magnetic, $\lambda =M$, multipoles, and $(T_{\lambda K})_0$ is the $0$ component of the operator $T_{\lambda K}$ in spherical coordinates, where $T_{E1} \equiv D$, $T_{M1} \equiv \mu$, and $T_{E2} \equiv Q_2$ 
%\MK{(for consistency with \eqref{Qlk} it is better to change $T_{\lambda k}$ to $T_{\lambda K}$)}. 
These many-electron operators are expressed by the sum of the single-electron operators. For example,
${\bm \mu} = \sum_{i=1}^N {\bm \mu}_i$, where $N$ is the number of electrons in the atom. The sum of $n$ in \eref{Qlk} includes the positive and negative energy states. In the following, we label the intermediate positive energy states by $n^+$ and the negative energy states by $n^-$.

In calculating the $E2$ polarizabilities, the contribution of intermediate negative energy states is completely negligible. The operator $Q_2 \sim r^2$ mixes the large components of the initial and final electronic wave functions in the matrix elements (MEs) $\langle n^+|Q_{20} |0 \rangle$. In a positron wave function, the large and small components are swapped and, respectively, $Q_2$ mixes the large and small components in ME $\langle n^-|Q_{20} |0 \rangle$. Since $r^2$ is a long-distance operator, it leads to suppression of ME
$\langle n^-|Q_{20} |0 \rangle$ by a factor of $1/c$ compared to
$\langle n^+|Q_{20} |0 \rangle$. Additionally, the contribution of
negative energy states is suppressed by large energy denominators. 
%\MK{Is it $1/c$, not $1/c^2$? The $1/c^2$ will appear for the matrix element squared.} 
For this reason, the results obtained in~\cite{PorSafSaf18} for the $E2$ polarizabilities remain valid, and we do not recalculate them here.

For the $M1$ polarizabilities of the clock states, the situation is quite different. The operator $M1$ in relativistic form is
${\bm \mu} = -1/2 \, ({\bm \alpha} \times {\bf r})$, where
$ {\bm \alpha} = \left(
 \begin{array}{ll}
      0        & {\bm \sigma} \\
  {\bm \sigma} &       0
 \end{array}
\right)$ and ${\bm \sigma}$ are the Pauli matrices. This operator mixes the large and small components of the wave functions in $\langle n^+|\mu_0 |0 \rangle$ and the large components of the wave functions in $\langle n^-|\mu_0 |0 \rangle$. Due to the presence of $\bf r$, the operator acts at long distances and, as a result, MEs $\langle n^+|\mu_0 |0 \rangle$ are suppressed by a factor of $1/c$ compared to $\langle n^-|\mu_0 |0 \rangle$. Note that this suppression is not compensated for by the large denominators $(E_0- E_{n^-})$ in the sum over $|n^-\rangle$ even for the $^3\!P_0^o$ state. Thus, the negative energy states give the dominant contribution to the $M1$ polarizabilities of both clock states at the magic frequency.

We note that the use of the relativistic form of the $M1$ operator is very important in correctly taking into account the contribution of negative-energy states. In the non-relativistic form
${\bm \mu} \sim ({\bf J} + {\bf S})$ and mixes only the large components of the wave functions. In our paper \cite{PorSafSaf18} we used the nonrelativistic form of this operator, which led to a significant underestimation of this contribution.

The formalism developed to calculate the $M1$ polarizabilities of the clock states is presented below. Using \eref{Qlk}, we can write the expression for the dynamic $M1$ polarizability of a $|0\rangle$ state as
\begin{eqnarray}
\alpha_{M1}(\omega) &=& 2
\left[\sum_{n=n^+} + \sum_{n=n^-}\right] \frac{\Delta E_n |\langle n |\mu_0|0 \rangle|^2}{(\Delta E_n)^2-(\omega)^2} ,
\label{aM1}
\end{eqnarray}
where $\Delta E_n \equiv E_n-E_0$.

The denominators in the second term of \eref{aM1} can be approximated by $E_{n^-}-E_0 = -2 c^2 [1+ O(1/c^2)] \approx -2 c^2 $. The typical values of the frequencies $\omega$, used in experiments, are much lower than
$2 c^2 \approx 3.8 \times 10^4$ a.u.. For example, the magic frequency
$\omega^* \approx 0.056\,\,{\rm a.u.}$. Neglecting $\omega$ compared to
$\Delta E_{n^-}$, we obtain
\begin{eqnarray}
2 \sum_{n^-} \frac{\Delta E_{n^-}|\langle n^- |\mu_0|0 \rangle|^2}{(\Delta E_{n^-})^2-\omega^2}
\approx - \frac{1}{c^2} \sum_{n^-} |\langle n^- |\mu_0|0 \rangle|^2 .
\label{a2}
\end{eqnarray}

Using this expression and also adding and subtracting to \eref{aM1} the similar term with summation over $n^+$,
$$-\frac{1}{c^2} \sum_{n^+} |\langle n^+ |\mu_0| 0 \rangle|^2
  +\frac{1}{c^2} \sum_{n^+} |\langle n^+ |\mu_0| 0 \rangle|^2,$$ we find
\begin{eqnarray}
\alpha_{M1}(\omega) &\approx& 2 \sum_{n^+} \left[\frac{\Delta E_{n^+}}{(\Delta E_{n^+})^2-\omega^2} + \frac{1}{2 c^2} \right]
                                |\langle n^+ |\mu_0| 0 \rangle|^2 \nonumber \\
                    &-& \frac{1}{c^2} \sum_n |\langle n |\mu_0|0 \rangle|^2 .
\label{aM1_1}
\end{eqnarray}

Now, the summation in the second term of \eref{aM1_1} goes over {\it all} intermediate states, and using the
%completeness condition
closure relation $\sum_n |n\rangle \langle n| =1$
%\textbf{[MK: This condition is usually addressed as closure]},
we can write
\begin{eqnarray}
\sum_n |\langle n |\mu_0|0 \rangle|^2 = \langle 0 | \mu_0^2| 0 \rangle
= \frac{1}{6}\, \langle 0 | r^2  |0 \rangle \left[1 + O\left(\frac{1}{c^2}\right)\right],
\label{sumr2}
\end{eqnarray}
where $r^2 \equiv \sum_{i=1}^N r_i^2$. The last expression in \eref{sumr2} was obtained after simple transformations using the properties of the matrices ${\bm \alpha}$ and ${\bm \sigma}$ and also the properties of a spherically symmetric state.

As we discussed in~\cite{PorSafSaf18}, only a few low-lying positive-energy intermediate states give dominant contributions to the polarizabilities $M1$, and it is sufficient to take them into account in the sum over $|n^+\rangle$. For these states
$$\left| \frac{\Delta E_{n^+}}{(\Delta E_{n^+})^2-(\omega^*)^2} \right| \gg \frac{1}{2 c^2}$$
and we can neglect $1/(2c^2)$ in the first term of \eref{aM1_1}.

In total, neglecting the terms $\sim 1/c^4$, we arrive at
\begin{eqnarray}
\alpha_{M1}(\omega) &\approx& 2 \sum_{n^+} \frac{\Delta E_{n^+}}{(\Delta E_{n^+})^2-\omega^2} |\langle n^+ |\mu_z| 0 \rangle|^2 \nonumber \\
                    &-& \frac{1}{6 c^2}\, \langle 0 | r^2  |0 \rangle .
\label{aM1_2}
\end{eqnarray}

The first term in \eref{aM1_2} is associated with the contribution of the positive-energy states and the second term with the contribution of negative-energy states.
In the following, we designate them by $\alpha^+_{M1}(\omega)$ and $\alpha^-_{M1}(\omega)$, respectively. Note that the second term is the same as the expression for the diamagnetic susceptibility of an atom given by the Langevin formula (see, e.g., Ref.~\cite{LanLif97}).

Let us briefly discuss the Breit correction to the first term in \eref{aM1_2}. The Breit operator includes $\bm \alpha$ matrices. Thus, when we calculate this correction to the ME $\langle n^+ |\mu_z| 0 \rangle$, we may also need to include the negative-energy state contribution. Note that the dominant Breit correction to the valence atomic states comes from the exchange with the innermost core state $1s$ \cite{KPT00a}. It is then easy to estimate that the negative-energy state contribution to the ME $\langle n^+ |\mu_z| 0 \rangle$ is on the order of $\alpha^3 Z$, where $\alpha Z$ comes from the small component of the $1s$ state. 
%This is larger than the naive expectation, but small enough to be neglected.

%=================================
\paragraph{Method of calculation.}
%=================================
We consider Sr as an atom with two valence electrons above the closed shell core and perform calculations within the framework of methods that combine configuration interaction (CI) with (i) many-body perturbation theory ~\cite{DzuFlaKoz96} and (ii) the linearized coupled-cluster method~\cite{SafKozJoh09}. In these methods, the energies and wave functions are found from the multiparticle Schr\"odinger equation
\begin{equation}
H_{\mathrm{eff}}(E_n) \Phi_n = E_n \Phi_n,
\label{Heff}
\end{equation}
where the effective Hamiltonian is defined as
\begin{equation}
H_{\mathrm{eff}}(E) = H_{\mathrm{FC}} + \Sigma(E).
\end{equation}
Here, $H_{\mathrm{FC}}$ is the Hamiltonian in the frozen core (Dirac-Hatree-Fock) approximation and $\Sigma$ is the energy-dependent correction, which takes into account virtual core excitations in the second order of the perturbation theory (the CI+MBPT method) or to all orders (the CI+all-order method).

The electric dipole polarizabilities of the Sr clock states were calculated at the magic frequency $\omega^*$ in Ref.~\cite{SafPorSaf13} to be
$\alpha_{E1}(\omega^*) = 286.0(3)$ a.u.. The $E2$ polarizabilities, as well as the contribution of the positive-energy states to the $M1$ polarizabilities of the clock states (given by the first term in \eref{aM1_2}) were calculated in our previous work ~\cite{PorSafSaf18}, so we only need to compute $\alpha^-_{M1}$.
%==========================================
\paragraph{Calculation of $\alpha^-_{M1}$.}
%==========================================
The calculation of the contribution of the negative-energy states to the $M1$ polarizabilities is reduced to the determination of a matrix element
%We need to find the following.
\begin{equation}
\alpha^-_{M1} \equiv  -\frac{1}{6 c^2}\, \langle 0 | r^2  |0 \rangle, 
\label{am1_m}
\end{equation}
where  $|0 \rangle$ is either $^1\!S_0$ or $^3\!P^o_0$ state.
Since $r^2$ is the scalar operator, one needs to calculate the contribution of valence and core electrons to $\langle 0 | r^2  |0 \rangle$. Consequently, we can divide the ME $\langle 0 | r^2  |0 \rangle$ into the valence and core parts as
$$\langle 0 | r^2 |0 \rangle = \langle 0 |r^2|0 \rangle_v + \langle 0 | r^2 |0 \rangle_c.$$ In the single-electron approximation, the core contribution is given by
\begin{eqnarray}
 \langle 0 | r^2  |0 \rangle_c = \sum_{a=1}^{N_c} \langle a | r_a^2 | a \rangle ,
\label{r2core}
\end{eqnarray}
where $|a\rangle$ is the single-electron wave function of the $a$th core electron and $N_c$ is the number of core electrons.

To find the valence parts of the MEs $\langle ^1\!S_0 | r^2 | ^1\!S_0 \rangle$ and $\langle ^3\!P_0^o | r^2 | ^3\!P_0^o \rangle$ and estimate their uncertainties, we carried out the calculation using the CI + MBPT and CI+all-order methods. The results are presented in \tref{Tab:MEr2}.

We note that the correlation corrections to the expectation values of the operator $r^2$ arise from the correlation corrections to the wave functions and the corrections to the operator. The latter include the random phase approximation (RPA), the two-particle and core Brueckner~\cite{DzuKozPor98}, and the structural radiation~\cite{DzuFlaSil87,BlaJohLiu89} and normalization corrections~\cite{DzuFlaKoz96}. All of them are small (for example, the RPA correction is less than 1\% for both MEs). In addition to that, these corrections to the operator tend to cancel each other out and, in total, give a very small contribution $\sim -0.1$~a.u.. It is given in the row ``$\Delta$ (Corrections)''. The core contribution was calculated using \eref{r2core} and is given in the row labeled ``Core''.
Total values were obtained as the sum of the CI+all-order value, ``$\Delta$ (Corrections)'' and the core contribution.

The uncertainty in the correlation correction for the wave function is estimated as the difference between the CI+all-order and CI+MBPT values, which is less  than 1.5\% (see~\tref{Tab:MEr2}).
However, this difference is positive for $^3\!P_0^o$ and negative for $^1\!S_0$. As a result, for $\Delta r^2 \equiv \langle ^3\!P_0^o | r^2 | ^3\!P_0^o \rangle$ - $\langle ^1\!S_0 | r^2 | ^1\!S_0 \rangle$, given in the last column of \tref{Tab:MEr2}, the difference between the CI+MBPT and CI+all-order values increases to 7-8\%.

In contrast, the core contribution to these MEs is large, amounting to 50-60\% of the valence contribution. The accuracy of the single-electron approximation, \eref{r2core}, is not very high. But the core contribution is the same for both MEs.
As a result, the total value of $\Delta r^2$ is determined by the difference in valence contributions because the core contributions cancel out. We estimate its uncertainty as the difference between the CI+MBPT and CI+all-order values.
% ####################################################################################################################
\begin{table}[h!]
\caption{Matrix elements $\langle ^1\!S_0 | r^2 | ^1\!S_0 \rangle$ and $\langle ^3\!P_0^o | r^2 | ^3\!P_0^o \rangle$, obtained
in the CI+MBPT and CI+all-order approximations, are given in a.u.. The sum of all corrections to the operator $r^2$,
described in the text, is given in the row ``$\Delta$ (Corrections)''.
The core contribution is given in the row ``Core''. The ``Total'' value = ``CI+all-order'' + $\Delta$ (Corrections) + ``Core''.
The values of $\Delta r^2$ are given in the column labeled ``$\Delta r^2$''. The uncertainty is given in parentheses.}
\label{Tab:MEr2}%
\begin{ruledtabular}
\begin{tabular}{cccc}
                        & $\langle ^1\!S_0 | r^2 | ^1\!S_0 \rangle$ & $\langle ^3\!P_0^o | r^2 | ^3\!P_0^o \rangle$ & $\Delta r^2$ \\
\hline \\ [-0.5pc]
% CI                     &           46.77                           &              57.26                           &  10.5 \\[0.3pc]
CI+MBPT                 &           43.3                            &              54.1                             &  10.8 \\[0.3pc]
CI+all-order            &           42.7                            &              54.3                             &  11.6 \\[0.5pc]
$\Delta$ (Corrections)  &           -0.1                            &              -0.1                             &       \\[0.5pc]
 Core                   &           26.4                            &              26.4                             &       \\[0.5pc]

 Total                  &           69.0                            &              80.6                             &  $11.6(8)$
\end{tabular}
\end{ruledtabular}
\end{table}
% ####################################################################################################################

The final values of $\alpha^{\pm}_{M1}$, $\alpha_{M1} = \alpha^{+}_{M1} + \alpha^{-}_{M1}$, and $\alpha_{E2} \approx \alpha^{+}_{E2}$ for the $^1\!S_0$ and $^3\!P_0^o$ states are presented in Table~\ref{Tab:alpha}. Quantities $\alpha^{-}_{M1}$ were obtained in this work, while $\alpha^{+}_{M1}$ and $\alpha_{E2}$ were taken from Ref.~\cite{PorSafSaf18}.

Comparing our results with those obtained in Ref.~\cite{WuShiNi23}, we see
a good agreement for all quantities except $\alpha^{-}_{M1}$. The valence contribution to $\alpha^{-}_{M1}$ also agrees very well with that obtained in Ref.~\cite{WuShiNi23} for both clock states. The difference from~\cite{WuShiNi23} in the total values of $\alpha^-_{M1}(^1\!S_0)$ and
$\alpha^-_{M1}(^3\!P_0^o)$ is due to the core contribution. We assume that the authors of Ref.~\cite{WuShiNi23} did not take it into account.

We find that $\alpha^+_{M1}(^1\!S_0)$ is negligible compared to $\alpha^-_{M1}(^1\!S_0)$. For the $^3\!P^o_0$ state, $\alpha^-_{M1}(^3\!P^o_0)$ is two orders of magnitude larger in absolute value than $\alpha^+_{M1}(^3\!P^o_0)$. Thus, the differential polarizability $M1$ $\Delta \alpha_{M1}$ is mainly determined by
the contributions of the negative energy states. The uncertainty 7\% of $\Delta \alpha_{M1}$ corresponds to the uncertainty of $\Delta r^2$. Using $\Delta \alpha_{M1}$ and $\Delta \alpha_{E2}$ we found $\Delta \alpha_{qm}$. Its absolute uncertainty was obtained as
$$ \Delta \alpha_{qm} = \sqrt{(\Delta \alpha_{M1})^2 + (\Delta \alpha_{E2})^2}.$$
%Our value of $\Delta \alpha_{qm}$ is in good agreement with the result obtained in %Ref.~\cite{WuShiNi23}.
% ##########################################################################################
\begin{table}[tbp]
\caption{The $M1$, $E2$, the differential polarizabilities for the $^1\!S_0$ and $^3\!P_0^o$ states and $\Delta \alpha_{qm}$ are presented (in a.u.). The values of $\alpha^+_{M1}$ and $\alpha_{E2}$ are from Ref.~\cite{PorSafSaf18}. Our values are compared with the results of Ref.~\cite{WuShiNi23}. The uncertainties are given in parentheses.}
\label{Tab:alpha}%
\begin{ruledtabular}
\begin{tabular}{lrr}
Polariz.                     & This work \& Ref.~\cite{PorSafSaf18}  &  Ref.~\cite{WuShiNi23} \\
\hline \\ [-0.5pc]
$\alpha^+_{M1}(^1\!S_0)$     &   $ 2 \times 10^{-9}$                  & $2.17 \times 10^{-9}$      \\[0.3pc]
$\alpha^-_{M1}(^1\!S_0)$     &   $-6.13 \times 10^{-4}$               & $-3.84 \times 10^{-4}$     \\[0.3pc]
$\alpha_{M1}(^1\!S_0)$       &   $-6.13 \times 10^{-4}$               & $-3.84(24) \times 10^{-4}$ \\[0.5pc]

$\alpha^+_{M1}(^3\!P^o_0)$   &   $-0.05 \times 10^{-4}$               & $-0.05 \times 10^{-4}$     \\[0.3pc]
$\alpha^-_{M1}(^3\!P_0^o)$   &   $-7.15 \times 10^{-4}$               & $-4.88 \times 10^{-4}$     \\[0.3pc]
$\alpha_{M1}(^3\!P^o_0)$     &   $-7.20 \times 10^{-4}$               & $-4.93(30) \times 10^{-4}$ \\[0.5pc]

$\Delta \alpha_{M1}$         &   $-1.07(7) \times 10^{-4}$            & $-1.09(38) \times 10^{-4}$ \\[0.5pc]

$\alpha_{E2}(^1\!S_0)$       &   $ 0.89(3) \times 10^{-4}$            & $ 0.928(57) \times 10^{-4}$     \\[0.3pc]
$\alpha_{E2}(^3\!P^o_0)$     &   $ 1.22(3) \times 10^{-4}$            & $ 1.244(76) \times 10^{-4}$     \\[0.5pc]

$\Delta \alpha_{E2}$         &   $ 0.33(4) \times 10^{-4}$            & $ 0.316(95) \times 10^{-4}$  \\[0.5pc]

$\Delta \alpha_{qm}$         &   $-0.74(8) \times 10^{-4}$            & $-0.77(39) \times 10^{-4}$
\end{tabular}
\end{ruledtabular}
\end{table}
% ##########################################################################################

% -----------------------------------------------
\paragraph{Comparison with experimental results.}
% -----------------------------------------------
In the experimental works of RIKEN~\cite{UshTakKat18}, PTB~\cite{DorKloPal23}, and JILA~\cite{KimAepBot23}
%it was determined the quantity $\tilde \alpha_{qm}/h$, expressed in Hz and connected with $\Delta \alpha_{qm}$ by the formula
the following quantity was measured (in Hz):
\begin{equation}
\frac{\tilde \alpha_{qm}}{h} \equiv \frac{\Delta \alpha_{qm}(\omega^*)}{\alpha_{\rm E1}(\omega^*)}\,\frac{E_r}{h} ,
\label{}
\end{equation}
where $E_{r}$ is the photon recoil energy and $h$ is the Planck constant. For $\lambda^* \approx 813.428$ nm we have
$$\frac{E_{r}}{h} = \frac{h}{2M\lambda^{*2}} \approx 3.47\,\, {\rm KHz},$$
where $M$ is the mass of the $^{87}$Sr atom.

Using our calculated value of $\Delta \alpha_{qm}$ and
$\alpha_{\rm E1}(\omega^*)=286.0(3)$ a.u.~\cite{SafPorSaf13} we find
$\tilde \alpha_{qm}/h$ and compare it with other results in \tref{Tab:alphaqm}.
%The theoretical value cited in the table as the result of the Wuhan group %was obtained using their value of $\Delta \alpha_{qm}$ and our value of %$\alpha_{\rm E1}$. 
% ####################################################################################################################
\begin{table} %[th!]
\caption{The values of $\tilde \alpha_{qm}/h$ (in mHz) and $\Delta \alpha_{qm}$ (in a.u.), found at the magic frequency,
are presented.}
\label{Tab:alphaqm}%
\begin{ruledtabular}
\begin{tabular}{clcc}
           &                               &   $\tilde \alpha_{qm}/h$ &    $\Delta \alpha_{qm}$ \\
\hline \\ [-0.5pc]
  Theory   &  This work:                   &     $-0.90(10)$          &   $-7.4(8)  \times 10^{-5}$    \\[0.2pc]
           & Wuhan~\cite{WuShiNi23}       &     $-0.94(48)$          &   $-7.7(3.9) \times 10^{-5}$   \\[0.5pc]

Experiment & JILA 2022~\cite{KimAepBot23}  &     $-1.24(5)$           &                                \\[0.2pc]
           & RIKEN~\cite{UshTakKat18}      &     $-0.96(4)$           &                                \\[0.2pc]
           & PTB~\cite{DorKloPal23}        &     $-0.99(20)$          &
\end{tabular}
\end{ruledtabular}
\end{table}
% ####################################################################################################################

To conclude, we derived an expression for the $M1$ polarizability that accounts for the contribution of the positive- and negative-energy states. To calculate $\alpha_{M1}^-$ we used a simple but accurate formula given by \eref{am1_m}. Using this formula we found the contribution of the negative-energy states to the $M1$ polarizabilities of the $^1\!S_0$ and $^3\!P_0^o$ states at the magic frequency and showed that this contribution is completely dominant for both clock states. Given the new values of the $M1$ polarizabilities and the values of the $E2$ polarizabilities obtained
in Ref.~\cite{PorSafSaf18}, we found the quantities $\Delta \alpha_{qm}$ and
$\tilde \alpha_{qm}/h$. Comparing the latter with the experimental results, we observe good agreement between the theory and the experiment, resolving the contradiction between the theoretical and experimental results.

We thank Joonseok Hur, Kyungtae Kim, Ilya Tupitsyn, Will Warfield, and Jun Ye for fruitful discussions.
This work was supported in part by NSF QLCI Award OMA - 2016244, ONR Grant No. N00014-20-1-2513, and by the European Research Council (ERC) under the European Union Horizon 2020 research and innovation program (Grant No.~856415).

%\bibliography{Sr_paper}

\end{document}